\documentclass[prb,aps,twocolumn,amsmath,amssymb,floatfix,superscriptaddress]{revtex4-1}
\usepackage{graphicx,amsmath,bm}
\usepackage{float}
\usepackage{subfigure}
\usepackage{mathrsfs}
\usepackage{graphics}
\usepackage{epstopdf}
\usepackage{amssymb}
\usepackage{hyperref}
\usepackage{color}
\usepackage{braket}
\usepackage{multirow}
\begin{document}

\author{Kallol Mondal}
\email[E-mail: ]{kallolsankarmondal@gmail.com}
\affiliation{School of Physical Sciences, National Institute of Science Education and Research Bhubaneswar, HBNI, Jatni, Odisha-752050, India}

\author{Sudin Ganguly}
\email[E-mail: ]{sudinganguly@gmail.com}
\affiliation{Department of Physics, School of Applied Sciences, University of Science and Technology Meghalaya, Ri-Bhoi-793101, India }

\author{Santanu K. Maiti}
\email[E-mail: ]{santanu.maiti@isical.ac.in}
\affiliation{Physics and Applied Mathematics Unit, Indian Statistical Institute, 203 Barrackpore Trunk Road, Kolkata-700108, India}

\date{\today}
\title{Thermoelectric phenomena in an antiferromagnetic helix: Role of electric field}
\begin{abstract}
The charge and spin-dependent thermoelectric responses are investigated on a single-helical molecule possessing a collinear antiferromagnetic spin arrangement with zero net magnetization in the presence of a transverse electric field. Both the short and long-range hopping scenarios are considered, which mimic biological systems like single-stranded DNA and $\alpha$-protein molecules. A non-equilibrium Green's function formalism is employed following the Landauer-Buttiker prescription to study the thermoelectric phenomena. The detailed dependence of the basic thermoelectric quantities on helicity, electric field, temperature etc., are elaborated on, and the underlying physics is explained accordingly. The charge and spin \textit{figure of merits} are computed and compared critically. For a more accurate estimation, the phononic contribution towards thermal conductance is also included. The present proposition shows a favorable spin-dependent thermoelectric response compared to the charge counterpart.
\end{abstract}
\maketitle
\section{\label{sec:intro}Introduction}

Achieving a favorable thermoelectric (TE) response is a long-sought goal in the material science community to overcome the dilemma of the global energy crisis. This is due to the fact that heat-to-energy conversion potentially can be an effective mechanism for scavenging waste heat~\cite{snyder2008,He-tritt-Science} by developing efficient devices. Even after persistent efforts and investments, designing efficient thermoelectrics is reaching a plateau. The obtained efficiency is not up to the mark and hence, is far from commercialization. The efficiency of TE material is characterized by a dimensionless parameter, namely \textit{figure of merit} (FOM), denoted by $ZT$, which explicitly depends on the Seebeck coefficient, electrical conductance, temperature, and total thermal conductance~\cite{PhysRevB.47.12727}. For bulk systems, electrical and thermal conductances are correlated by the Wiedemann-Franz (W-F) law~\cite{doi:10.1126/science.1093164}, which essentially restricts to have an efficient energy conversion. However, it is possible to achieve better TE performance in the nanoscale regime than the bulk ones, overshadowing the W-F law~\cite{PhysRevLett.100.066801,PhysRevB.47.12727,PhysRevB.47.16631}. Extensive efforts have been made to study the thermoelectric phenomena exploring the charge degrees of freedom in the nanoscale regime with systems like quantum dots~\cite{PhysRevB.79.081302,PhysRevB.80.195409,PhysRevB.81.205321,PhysRevB.84.075410,PhysRevB.85.085408}, nanowires~\cite{PhysRevB.84.205410,PhysRevB.86.035433,C3TA01594G,PhysRevLett.120.177703}, topological insulators~\cite{PhysRevLett.112.136402,Xu2017}, and also organic molecular junctions~\cite{PhysRevB.72.165426,doi:10.1126/science.1137149,PhysRevB.81.235406,RevModPhys.83.131} including DNAs, protiens~\cite{Macia2005,PhysRevB.75.035130,PhysRevB.82.045431,DONG2015176,Li2016}, etc.

On the other hand, compared to charge-based devices, spintronic devices are usually faster, more efficient, and have smaller dimensions where the electron's spin allows us to perform more work providing much less effort~\cite{Nikonov2006,PhysRevLett.55.1790,PhysRevLett.61.2472,RevModPhys.76.323,PhysRevLett.81.1282}. A recent development in the field of thermoelectric has been the entry of spin degrees of freedom, and magnetic order provides a `green' strategy to enhance the thermoelectric figure of merit~\cite{Zheng_2012}. This is due to the fact that the TE efficiency is directly proportional to the square of the Seebeck coefficient~\cite{goldsmid2010introduction}, and for a spin TE, it is defined as the difference between the contributions from the up and down spins. Interestingly, for the spin TE case, it is possible to achieve different signs of the spin-Seebeck coefficient, which can add up to produce a favorable TE response. Precisely, the spin-Seebeck effect(SSE)~\cite{PhysRevB.35.4959,Bauer2012} is the charge analog of the Seebeck effect, where one can generate a net spin current from the temperature gradient and can potentially reduce the thermal dissipation induced by the total charge current~\cite{saitoh2006conversion,Valenzuela2006,PhysRevLett.98.156601}. One of the remarkable features of the spin-Seebeck device is that it possesses a scalability different from that of usual charge-based Seebeck devices, where the output power is proportional to the length perpendicular to the temperature gradient. Not only that, the heat current and charge current follow separate paths in the spin-based Seebeck device compared to the charge-based Seebeck device, which prompts us to think about that the spin Seebeck device as a possible route to enhance the thermoelectric FOM~\cite{Adachi_2013}. These salient features have invigorated spintronic research to develop spin-based TE devices~\cite{Kirihara2012,Uchida_2011,Uchida_2012}.

The primary requirement of a spintronic device is to look for an efficient mechanism that sets apart the charge carriers based on their spin quantum number, which essentially means achieving polarized spin current from a completely unpolarized electron beam. Among several propositions~\cite{PhysRevB.98.075435,Gauthier2019,Tsukagoshi1999}, the most studied one is the use of ferromagnetic material as a functional element~\cite{Kamboj_2019}. However, there are several limitations to overcome in that case, like a large resistivity mismatch is induced across the junction formed by ferromagnetic and non-magnetic materials, which act against the flow of the injected electrons~\cite{Kamboj_2019,PhysRevB.62.R16267}. Another major issue is the tuning of spin-selective junction currents under the application of external magnetic fields. Experimentally, it is hard to achieve such strong confinement of magnetic fields within the quantum regime. Due to the above-mentioned limitations, in the recent past, the focus is shifted towards spin-orbit (SO) coupled systems instead of ferromagnetic materials. The investigation along the line is dominated by Rashba SO coupled system over the Dresselhaus one, as the strength of the former one can be tuned externally by suitable setups~\cite{premasiri2018tuning,ganguly2021selective}. Extensive efforts have been made in this regard to explore a range of different geometries using inorganic and organic molecules~\cite{PhysRevB.73.075303,PhysRevB.73.155325,dey2013spin}. But it turns out that, especially in molecular systems, the strength of the SO coupling is significantly weak compared to the hopping strength, differing by order of magnitude~\cite{Su_2016}. In addition to that, the tuning of SO coupling strength is also restricted by external means. As a result, it is difficult to obtain a high degree of spin separation and its possible tuning in a wide range in those spin-orbit coupled systems.

Due to the aforementioned issues with ferromagnetic systems, there is a growing inclination towards antiferromagnetic materials for future spintronic applications~\cite{RevModPhys.90.015005,Jungwirth2018,JUNGFLEISCH2018865}. Antiferromagnets are magnetically ordered, with the nearest-neighbor spins aligning in the opposite direction resulting in net zero magnetic moments. Thus, these types of magnetic structures are robust against external perturbations like magnetic fields, produce no stray fields, display ultrafast dynamics, and are capable of generating large magnetotransport effects~\cite{artemchuk2020terahertz}. Intensive efforts have been made to unravel the spin transport properties in antiferromagnetic materials, and antiferromagnetic spintronics remains an active area of cutting-edge research~\cite{doi:10.1126/science.aab1031,Nemec2018,Kosub2017,gupta2020can}.

Recent experiments have made significant progress along the line, including biological systems, finding that double-stranded DNA (dsDNA) molecules are highly efficient spin filters~\cite{Ghler2011}. The results are remarkable in the sense that the DNA molecules are nonmagnetic, and the present spin-orbit couplings (SOCs) are too small to host the chiral-induced spin selectivity (CISS) effect. Interestingly, this CISS effect led us to think about exploring chiral molecules in spintronic applications and may shed light on the spin effects in biological systems~\cite{PhysRevLett.108.218102,doi:10.1073/pnas.1407716111,doi:10.1073/pnas.1311493110,PhysRevB.85.081404,PhysRevB.92.115418,PhysRevB.95.155411}. Due to the above reasons, the biological systems like, double-stranded DNA, single-stranded DNA, $\alpha$-protein with helical geometries are of particular interest.

\begin{figure}[t!]
\centering
\includegraphics[width=0.46\textwidth]{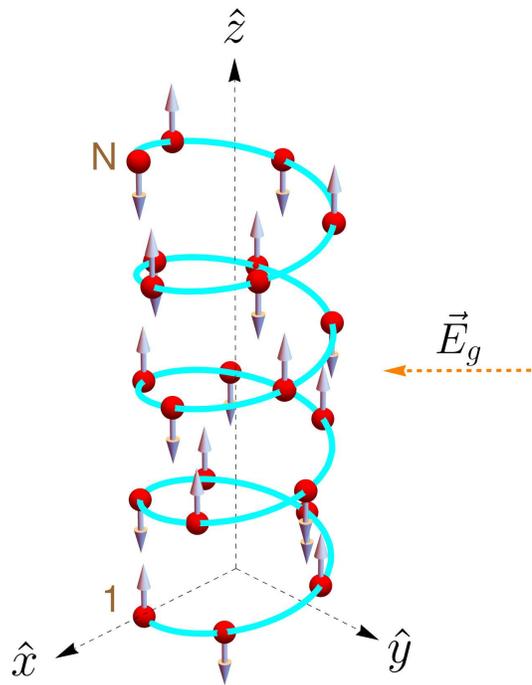} 
\caption{(Color online). Schematic diagram of an antiferromagnetic right-handed helix. Each red ball corresponds to a magnetic site and the arrow on the ball represents the direction of magnetic moment. Perpendicular to the helix axis, an external electric field is applied, which plays a central role in our investigation.}
\label{fig1:diagram}
\end{figure}

In the present communication, we propose a new prescription for efficient thermoelectric response, considering an antiferromagnetic helix as a functional element in the presence of the transverse electric field. To the best of our knowledge, no effort has been made to understand thermoelectric physics in such system, and this is precisely the driving force behind the present work. We extensively study the charge and spin-dependent thermoelectric responses on a single-stranded antiferromagnetic helix system connected by two one-dimensional (1D) non-magnetic, reflectionless, semi-infinite leads in the presence of a transverse electric field(see Fig.~\ref{fig1:diagram}). We simulate the whole system using the tight-binding framework. We employ non-equilibrium Green's function (NEGF) formalism following the Landauer-Buttiker prescription to study the thermoelectric phenomena~\cite{datta1997electronic,PhysRevB.23.6851,di2008electrical,nikolic2000quantum}. It is a well-known fact that no spin-separation is possible for antiferromagnetic systems with zero net magnetzation, {\it but it is possible to generate spin-filtration under the application of a transverse electric field}. The physics of spin-filtration gyrates around the interplay between the helicity of the antiferromagnetic helix (AFH) and the applied electric field. For a realistic estimation of the TE response, we also include the phonon contribution for the present case. To make this study complete, we also explore the thermoelectric responses across different temperatures. Our prescription shows a favorable spin-dependent thermoelectric response as compared to the charge counterpart at room temperature.

The rest part of the present communication is organized as follows. In Sec. II, we discuss the system along with the relevant interaction considered in the model and present the theoretical framework. All the results, considering the short-range and long-range interactions in the presence of an electric field, are critically investigated in Sec. III. Finally, in Sec. IV, we conclude our essential findings.

\section{\label{sec:formalism}Theoretical formulation}
\subsection{Description of the system}Let us first introduce the system to study the thermoelectric phenomena. Figure~\ref{fig1:diagram} depicts the schematic diagram of our proposed setup where a single-stranded antiferromagnetic helix possessing $N$ magnetic sites is attached to two 1D non-magnetic, reflectionless, semi-infinite leads, namely source ($S$) and drain ($D$) (not shown in the figure) at site $1$ and site $N$ respectively. These two leads are operating at two different temperatures, $T + \Delta T$ and $T- \Delta T$, where $T$ is the equilibrium temperature and $\Delta T$ is infinitesimally small. Thus, we restrict ourselves within the linear response regime throughout the analysis.

We use helical system as a functional element to study the TE response. In general, a helical system is described by two important parameters like stacking distance and twisting angle, denoted by $\Delta z$ and $\Delta \phi$ respectively~\cite{PhysRevLett.108.218102,doi:10.1073/pnas.1407716111}. These two parameters play a crucial role in determining whether the hopping is short-range or long-range and also determine the structure of the magnetic helix. When $\Delta z$ is very small, the atomic sites are closely spaced, and the electrons can hop to higher-order neighbor sites, yielding a long-range hopping (LRH) helix. On the other hand, when $\Delta z$ is quite large, the hopping of the electrons is restricted mostly to a few neighboring sites, and we have short-range hopping (SRH) helix. Here, we present the parameter values for a typical case of SRH and LRH in a tabular form in Table.~\ref{table}. (For the details of the helical geometry and relevant parameters, one may look at some previous pioneering efforts~\cite{PhysRevB.92.115418,PhysRevB.95.155411}.)
\begin{table}[h!]
\begin{center}
\begin{tabular}{ |c|c| c| c |c|c|}
\hline
\hline
 System & Radius & Stacking & Twisting & Decay\\
      type          &  (R)            &  distance $(\Delta z)$ & angle ($\Delta \phi$) &  constant $(l_c)$\\
\hline
 SRH & 7 \AA& 3.4 \AA & $ \pi/5$ rad& 0.9 \AA\\
 \hline
LRH & 2.5 \AA & 1.5 \AA & $5 \pi/9$   rad    & 0.9 \AA\\
 \hline
\end{tabular}
\label{table}
\caption{\label{table}Geometrical parameters for the helical sytem.}
\end{center}
\end{table}
\noindent These values mimic biological systems like single-stranded DNA and $\alpha$-protein molecules, and they are the most suitable examples where respectively, the short-range hopping and long-range hopping can be explored.

In our chosen antiferromagnetic helix system, the successive magnetic moments are aligned along $\pm z$ directions, and thus the resultant magnetization becomes zero. Each magnetic site $i$ is associated with a net spin $\langle \mathbf{S}_i\rangle$. The general orientation of any such spin vector can be described by the usual polar angle $\theta_i$ and the azimuthal angle $\phi_i$. Now, the incoming electron will interact with these local magnetic moments through the usual spin-moment exchange interaction $J$. To include this interaction, we introduce a spin-dependent scattering (SDS) parameter at each site $i$ as $\mathbf{h}_i = J \langle \mathbf{S}_i \rangle$~\cite{Su_2016}. The strength of the SDS parameter $|\mathbf{h}|$ is assumed to be isotropic, i.e., $\mathbf{h}_i = h ~~\forall ~i$. For the present investigation, the interaction between neighboring magnetic moments is ignored, and it is a subject of future study.

The central region i.e., the AFH, is exposed to an electric field, having strength $E_g$, perpendicular to the helix axis ($\hat{z}$) as shown in Fig.~\ref{fig1:diagram}. The incorporation of electric field in our theoretical formalism is described in the forthcoming sub-section.
\subsection{Model Hamiltonian}The  tight-binding Hamiltonian representing the total system comprises four parts,  which are given by~\cite{SHOKRI2005325,SHOKRI200653,DEY20101522,Patra2017}
\begin{equation}
\mathcal{H} = \mathcal{H}_\text{AFH}  + \mathcal{H}_\text{S}  + \mathcal{H}_\text{D}  +\mathcal{H}_\text{C}, 
\end{equation}
where,  $\mathcal{H}_\text{AFH}, \mathcal{H}_\text{S}, \mathcal{H}_\text{D},$  and $ \mathcal{H}_\text{C} $ represent the sub-parts of the Hamiltonian, associated with the AFH, source, drain, and the coupling between the leads and the AFH, respectively. 

The Hamiltonian for the AFH is given by~\cite{PhysRevB.95.155411,PhysRevB.100.205402}
\begin{eqnarray}
\mathcal{H}_\text{AFH} & = & \sum_{ n}  \mathbf{c}_n^\dagger \left(\bm{\epsilon}_n - \mathbf{h}_n \cdot \bm{\sigma} \right) \mathbf{c}_n \nonumber \\
& +& \sum_{n}^{N-1} \sum_m^{N-n} \left( \mathbf{c}_n^\dagger \bm{t}_{n} \mathbf{c}_{n+m} + h.c. \right),
\end{eqnarray} 
where $\mathbf{c}_n$ denotes the two-component fermioninc operator  at site $n$, given by $\mathbf{c}_n = \begin{pmatrix}c_{n \uparrow} \\c_{n \downarrow}\end{pmatrix}$ and its hermitian counterpart $\mathbf{c}^\dagger_n$ is  defined accordingly. $\bm{\sigma}$ is the well-known Pauli matrices, $\bm{t}_n$ and $\bm{\epsilon}_n$ are the $2 \times 2$ diagonal matrices given by
\begin{equation}
\bm{t}_n= \begin{pmatrix} t_n & 0\\ 0& t_n \end{pmatrix} ~~~~\text{and} ~~~ \bm{\epsilon}_n= \begin{pmatrix} \epsilon_n & 0\\ 0& \epsilon_n \end{pmatrix},
\end{equation}
where $\epsilon_n$ is the on-site energy in the absence of any spin-dependent scattering and $t_{n}$ represents the hopping amplitude from the site $n$ to $n+m$. The inclusion of SDS leads to the effective site energy matrix $\left(\bm{\epsilon}_n - \mathbf{h}_n \cdot \bm{\sigma}\right)$. Now, the presence of an external electric field $E_g$, perpendicular to the helix axis, modifies the on-site energy in the following way~\cite{PhysRevB.95.155411,PhysRevB.100.205402}
\begin{equation}
\epsilon_n^\text{eff} = \epsilon_n + e v_g \cos(n\Delta \phi - \beta),
\label{elec_ons}
\end{equation}
where $e$ is the electronic charge, $v_g\,(= 2E_g R)$ is the applied gate voltage, and $\beta$ is the angle between the incident electric field and the positive $\hat{x}$ axis, $R$ is the radius of the helix.

Due to the helical shape of the physical system, the hopping term becomes quite tricky, unlike the usual nearest-neighbor hopping (NNH) case. The summations over the site indices are to be taken carefully. The expression for the hopping integral $t_n$ is given by
\begin{equation}
t_n = t_1 \exp\left[-(l_n - l_1)/l_c\right],
\end{equation}
where $t_1$ and $l_1$ are the nearest-neighbor hopping amplitude and the distance among the nearest-neighbor sites, respectively.
$l_ c$ is the decay constant and $l_n$ is the spatial separation between the sites $n$ and $n+m$. The expression of $l_n$ is given by
\begin{equation}
l_n = \left[\left( 2R \sin\left(n \Delta \phi/2\right)\right)^2+ \left( n \Delta z \right)^2\right]^{1/2},
\end{equation}
where $\Delta z$ and $\Delta \phi$ are the stacking distance and twisting angle, respectively.
 
 The contributions from the leads and the coupling between the leads and the central region to the total Hamiltonian read as
 \begin{subequations}
 \begin{eqnarray}
 \mathcal{H}_{\rm S} & = & \sum_{m<1} \bm{a}_m^\dagger \bm{\epsilon}_0 \bm{a}_m +  \sum_{m<1}\left( \bm{a}_m^\dagger \bm{t}_0 \bm{a}_{m-1} + h.c. \right) \\
 \mathcal{H}_{\rm D} & = & \sum_{m>N} \bm{b}_m^\dagger \bm{\epsilon}_0 \bm{b}_m +  \sum_{m>N}\left( \bm{b}_m^\dagger \bm{t}_0 \bm{b}_{m+1} + h.c. \right) \\
  \mathcal{H}_{\rm C} & = & \bm{a}_0^\dagger \bm{\tau}_{\rm S} \bm{c}_1 +  \bm{c}_N^\dagger \bm{\tau}_{\rm D} \bm{b}_{N+1} + h.c.
 \end{eqnarray}
 \end{subequations}
Here, $\bm{a}_n,\bm{b}_n$ are used for the source and the drain in the same way like the $\bm{c}_n$ operator. $\bm{\epsilon}_0$ and $\bm{t}_0$ are $2 \times 2$ diagonal matrices where the on-site potential $\epsilon_0$ and hopping amplitude $t_0$  are taken to be the same for both the leads. The coupling between the source (drain) and the AFH is denoted by $\bm \tau_S\,(\bm \tau_{D})$, defined in the same footing as $\bm{t}_0$. 

\subsection{Two-terminal transmission probability}
We employ NEGF formalism to evaluate the two-terminal transmission probability through the helix system. The standard way to put up the retarded Green's function for the present case is as follows,
\begin{equation}
\mathcal{G}^r = \left[(E+ i ~0 ^+)\mathbb{I}- \mathcal{H}_{\text{AFH}}- \Sigma_{\sigma S} -\Sigma_{\sigma D}\right]^{-1},
\end{equation}
where $\sigma, \sigma^\prime$ are the spin indices, $\Sigma_{\sigma S}$ and $ \Sigma_{\sigma D}$ represent the contact self-energies of the source and drain, respectively, $\mathbb{I}$ is the identity matrix with dimension $2N \times 2N$. The rest of the other symbols have the usual meaning. 

Now, the transmission probability can be expressed in terms of retarded ($\mathcal{G}^r$) and advanced $\left(\mathcal{G}^a\left(=\mathcal{G}^r\right)^\dagger\right)$ Green's functions as
\begin{equation}
\mathcal{T}_{\sigma \sigma^\prime}= \text{Tr}\left[\Gamma_{\sigma S}~ \mathcal{G}^r~ \Gamma_{\sigma^\prime D} ~\mathcal{G}^a \right],
\label{eq:trans1}
\end{equation}
where $\Gamma_{\sigma S}$ and $\Gamma_{\sigma D}$ are the coupling matrices that describe the rate at which particles scatter between the leads and the AFH. $\mathcal{T}_{\sigma \sigma^\prime}$ indicates the probability of a transmitted electron with spin $\sigma^\prime$ injected with spin $\sigma$. We must mention that if $\sigma = \sigma^\prime$, then we get pure spin transmission, otherwise we get a spin-flip transmission. We define the net up and down spin transmission probabilities as
\begin{eqnarray}
\mathcal{T}_\sigma & = & \sum_{\sigma^\prime}\mathcal{T}_{\sigma^\prime \sigma} ,
\end{eqnarray}
where $\sigma, \sigma^\prime$ can be either $\uparrow$ or $\downarrow$. These are fundamental entities to calculate different thermoelectric quantities as described in the next sub-section.

\begin{figure*}[ht!]
\centering
\includegraphics[width=0.45\textwidth]{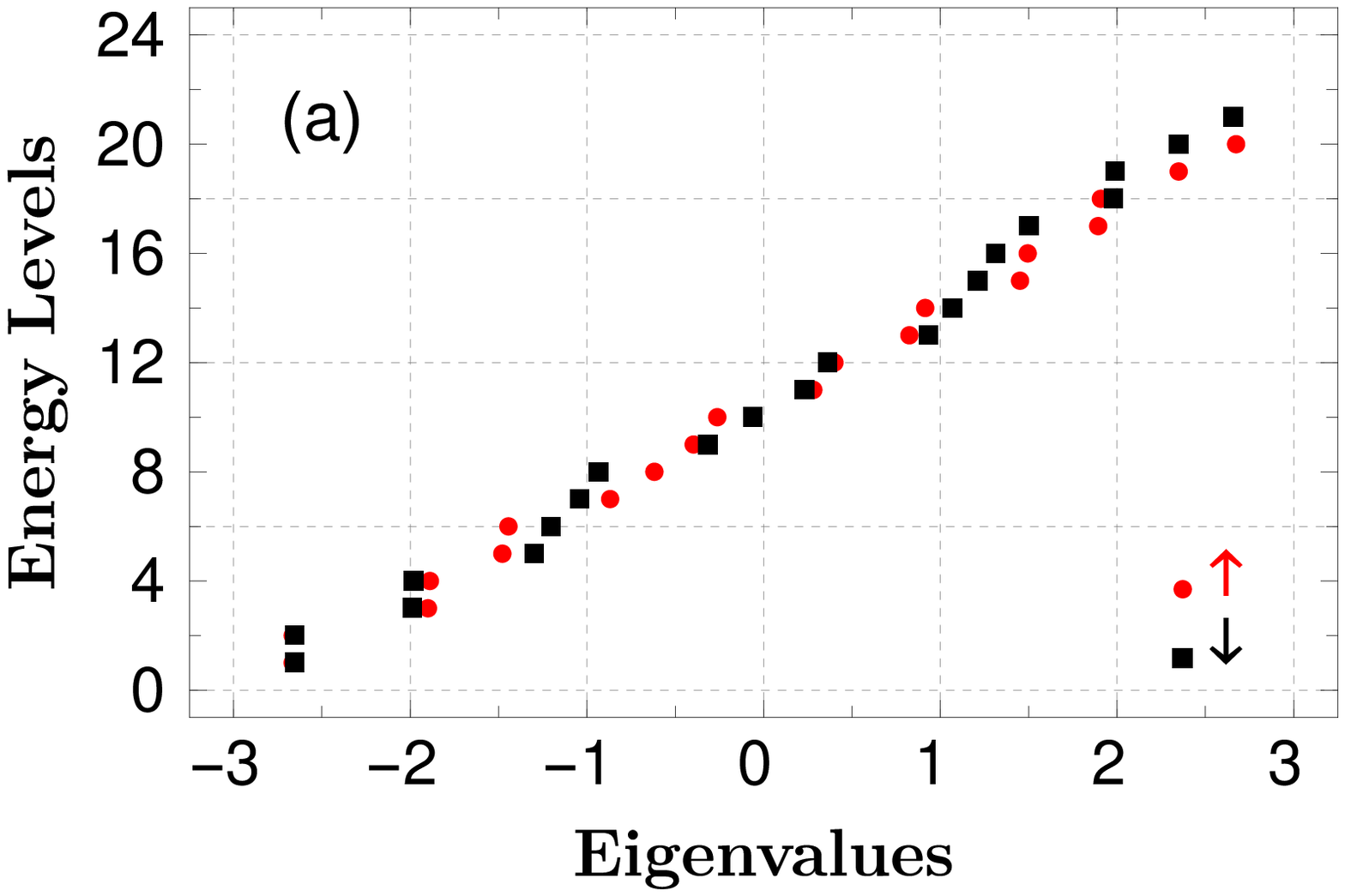} \hspace{0.2cm}
\includegraphics[width=0.45\textwidth]{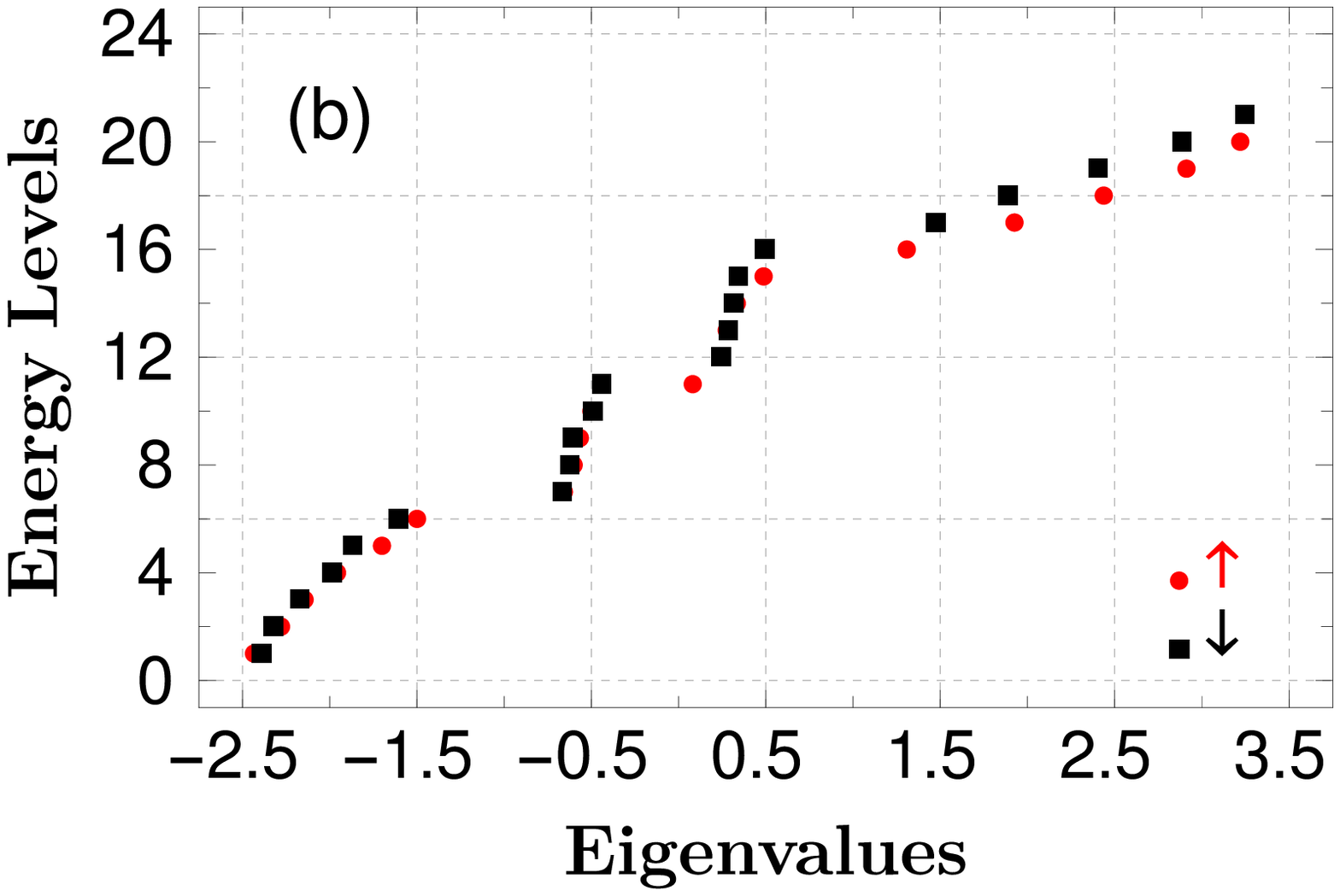} 
\caption{(Color online). Eigenvalue spectrum in presence of the electric field in case of (a) short-range hopping and (b) long-range hopping with $h=0.5$, $v_g =1$, and $\beta=0$. Number of sites in the Helix is $N=20$. The eigenvalues for the up and down spins are represented by red and black colors, respectively. Along $y$-direction, different energy levels are shown.}
\label{fig:LRH}
\end{figure*}

\subsection{Thermoelectric quantities}  
In the linear response regime, all the spin-resolved thermoelectric quantities like $G_\sigma$, $S_\sigma$, and $k_{\sigma\text{el}}$ can be extracted using Landauer’s integrals as~\cite{PhysRevApplied.3.064017,PhysRevB.79.033405}

\begin{subequations}
 \begin{eqnarray}
 G_\sigma & = & \frac{e^2}{h} L_{0 \sigma}, \\
 S_\sigma & = & -\frac{1}{e T }\frac{L_{1\sigma}}{ L_{0\sigma}},\\
k_{\sigma \text{el}} & = & \frac{1}{hT}\left( L_{2\sigma} - \frac{L_{1\sigma}^2}{L_{0\sigma}}\right),
 \end{eqnarray}
  \label{eq:TE-quantity}
\end{subequations}
where spin-resolved Landauer’s integral is given by
\begin{equation}
L_{n\sigma} = - \int   \mathcal{T}_\sigma(E) (E- E_F)^n\frac{\partial f_{\text{FD}}}{\partial E} ~dE,
\end{equation}
where, $h,f_{\text{FD}}$, and $E_F$ denote Planck’s constant, equilibrium Fermi-Dirac occupation probability, and Fermi energy, respectively. Here, $\mathcal{T}_\sigma(E)$ is the spin-resolved two-terminal transmission probability as defined earlier. 

Now, we define the charge ($c$) and spin ($s$) electrical conductances in the following way~\cite{PhysRevB.97.235421}
\begin{equation}
  G_{c}  =  G_\uparrow +  G_\downarrow  ~~~ \text{and}~~~  G_{ s} = G_\uparrow -  G_\downarrow .
\end{equation}

The charge and spin Seebeck coefficients (thermopowers) are defined by~\cite{PhysRevB.97.235421,PhysRevB.103.115424}
 \begin{equation}
 S_c  =  \frac{1}{2}\left(S_\uparrow + S_\downarrow \right) ~~~\text{and}~~~S_s  =  \left(S_\uparrow - S_\downarrow \right).
\end{equation}

Similarly, the charge and spin thermal conductances are given by~\cite{PhysRevB.97.235421}
 \begin{eqnarray}
 k_{c \text{el}}  =  k_{s \text{el}}= \left(k_\uparrow + k_\downarrow \right).
\end{eqnarray}

 The charge and spin \textit{figure of merits} can be expressed in a compressed form in the following way~\cite{PhysRevB.97.235421}
\begin{equation}
Z_\alpha T = \frac{\lvert G_\alpha\rvert S_{\alpha}^2 ~T}{k_\alpha (=k_{\alpha\text{el}}+ k_\text{ph})},
\end{equation}
where $\alpha \,(=\text{c, s})$ stands for the charge and spin degrees of freedom, $k_\text{ph}$ is the phonon contribution to the total thermal conductance and $T$ is the equilibrium temperature. Typically, a thermoelectric response of the order of unity is often regarded as favorable TE response. However, for an economically competitive response, $Z_\alpha T \sim 3$ is often prescribed~\cite{Tritt-review}. For a precise estimation of $Z_\alpha T$, one needs to consider the contribution of $k_\text{ph}$ in thermal conductance. The method for calculating $k_\text{ph}$ is given in the forthcoming sub-section.

\subsection{Calculation of phonon thermal conductance} 
When the temperature difference between the two contact electrodes is infinitesimally small, the phonon thermal conductance in the NEGF formalism can be evaluated from the expression~\cite{zhang-ph,hopkins-prb,aghosh,Mondal2021}
\begin{equation}
k_{\text{ph}}= \frac{\hslash}{2\pi}\int_0^{\omega_c} \mathcal{T}_{\text{ph}}\frac{\partial f_{BE}}{\partial T}\omega d\omega .
\end{equation}
Here, $\omega$ is the phonon frequency and $\omega_c$ the phonon cut-off frequency respectively. We consider only elastic scattering in the present case. $f_{BE}$ denotes the Bose-Einstein distribution function. $\mathcal{T}_{\text{ph}}$ is the phonon transmission probability across the central region, evaluated through the NEGF formalism as
\begin{equation}
\mathcal{T}_{\text{ph}}= \text{Tr}\left[\Gamma_S^{\text{ph}} \mathcal{G}_{\text{ph}} \Gamma_D^{\text{ph}} \left(\mathcal{G}_{\text{ph}}\right)^\dagger \right]
\end{equation}
$\Gamma_{S/D}^{\text{ph}}=i\left[\widetilde{\Sigma}_{S/D}-\widetilde{\Sigma}_{S/D}^\dagger\right]$ is known as the thermal broadening. $\widetilde{\Sigma}_{S/D}$ is the self-energy matrix for the source/drain electrode. The phononic Green's function for the AFH reads as
\begin{equation}
\mathcal{G}_{\text{ph}} = \left[{\mathbb M}\omega^2 - {\mathbb K} -\widetilde{\Sigma}_S - \widetilde{\Sigma}_D\right]
\end{equation}

where ${\mathbb M}$ is a diagonal matrix that describes the mass matrix of the helix. Each element of the mass matrix ${\mathbb M}_{nn}$ denotes the mass of the $n$-th atom in the helical system and ${\mathbb K}$ is the matrix of spring constants. The diagonal element ${\mathbb K}_{nn}$ denotes the restoring force of the $n$-th atom due to its neighboring atoms, while the element ${\mathbb K}_{nm}$ represents the effective spring constant between $n$-th and $m$-th neighboring atoms. The self-energy matrices $\widetilde{\Sigma}_S$ and $\widetilde{\Sigma}_D$ have the same dimension as ${\mathbb M}$ and ${\mathbb K}$ and can be computed by evaluating the self-energy term $\Sigma_{S/D}=-K_{S/D}\,\text{exp}\left[2i\,\text{sin}^{-1}\left(\frac{\omega}{\omega_c}\right)\right]$, where $K_{S/D}$ is the spring constant at the electrode-helix contact interface.

The spring constants are determined from the second derivative of Harrison's interatomic potential~\cite{harrison}. Since a 1D system does not allow any transverse interaction~\cite{kittel}, the spring constant for the 1D electrode is given by $K=3dc_{11}/16$. For a 3D system like helix, the spring constant is $K=3d\left(c_{11}+2c_{12}\right)/16$. Here $d$ denotes the interatomic spacing and $c_{11}$ and $c_{12}$ are the elastic constants. The cut-off frequency for the 1D electrode is determined from the relation $\omega_c=2\sqrt{K/M}$, in terms of the mass and spring constant.

\section{Numerical results and discussion}
\begin{figure*}[t!]
\centering
\includegraphics[width=0.475\textwidth]{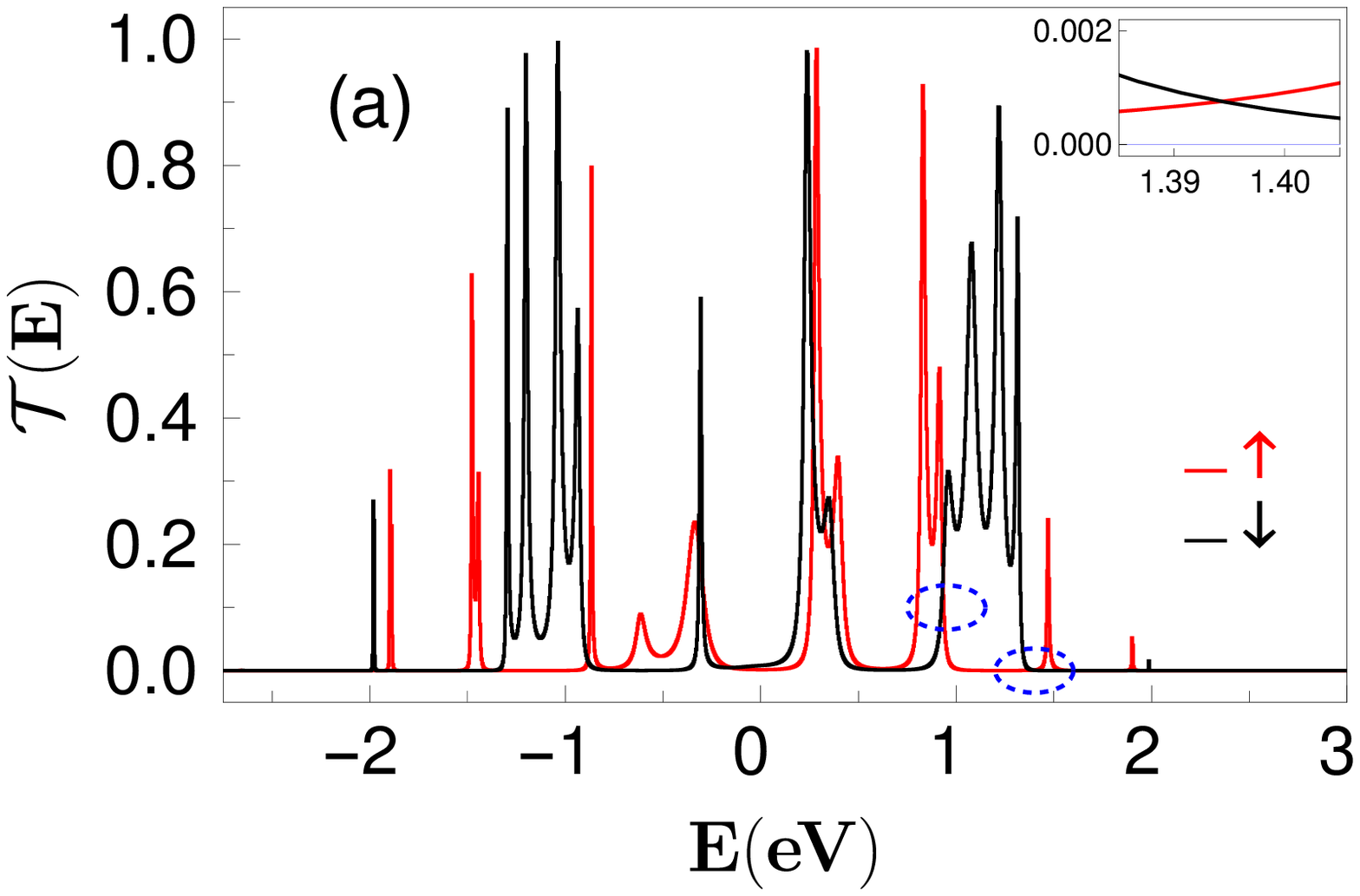} \hspace{0.2cm}
\includegraphics[width=0.475\textwidth]{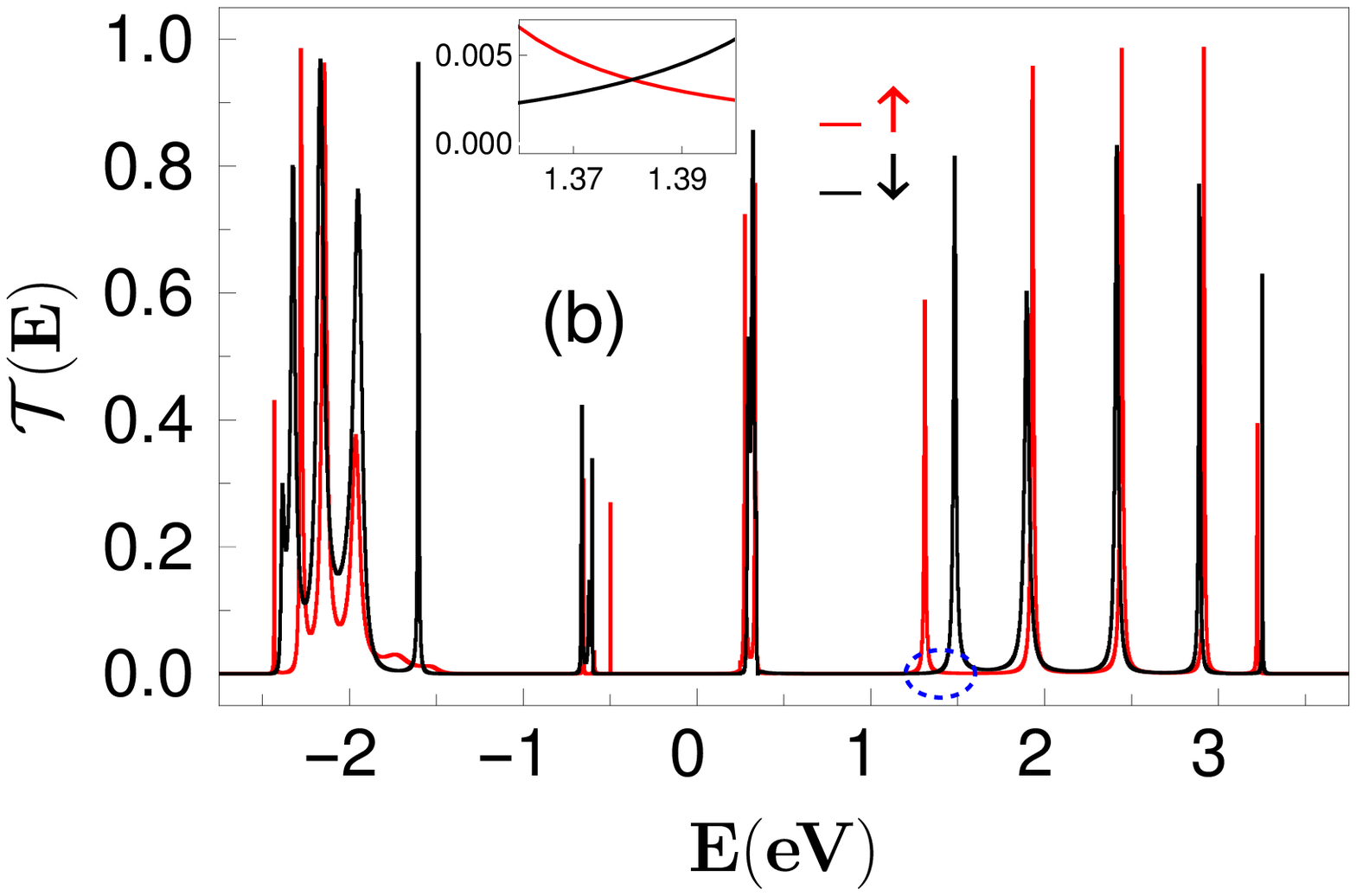} 
\caption{(Color online). Spin-resolved transmission probability as a function of energy for (a) short-range hopping and (b) long-range hopping. All the parameters and color conventions are the same as described in Fig.~\ref{fig:LRH}. The blue dotted ellipses mark the cross-over regions between up and down spin transmissions. In the insets, the cross-over is more visible and also shows that the transmission probabilities are small but finite.}
\label{fig:3}
\end{figure*}

The interplay between the transverse electric field and the helicity plays the central role to have spin-dependent TE phenomena in our chosen antiferromagnetic helix which we are going to discuss in this section. In the absence of any of these two parameters viz, the helicity and the electric field, there will be no mismatch between up and down spin channels, and therefore, we cannot expect any spin-dependent transport phenomena. The underlying physical mechanism is as follows. As all the magnetic moments are aligned along $\pm \hat{z}$ directions, the Hamiltonian of the antiferromagnetic helix can be decoupled as a sum of up and down spin Hamiltonians (viz, $H_\uparrow+H_\downarrow$). In the absence of electric field, these two sub-Hamiltonians are symmetric to each other, because of the anti-parallel configuration of the successive magnetic moments, resulting identical set of energy eigenvalues. The symmetry can be broken quite easily by applying an electric field in the helix system. Under that condition, we have a finite mismatch between the two different spin-dependent energy channels.

In presence of transverse electric field, the site energies get modulated in a cosine form as mentioned in Eq.~\ref{elec_ons}. The site energy expression looks identical to the well-known Aubrey-Andr\'{e}-Harper (AAH) model~\cite{aubry1980analyticity}. In the AAH model, the on-site term reads as $\epsilon_n=W\cos{(2\pi b n + \phi_\nu)}$ ($n$ being the site index), where $W$ is the AAH modulation strength, $b$ is an irrational number, and $\phi_\nu$ is the AAH phase. The one-to-one mapping is obvious in view of Eq.~\ref{elec_ons}, where one identifies the term $e v_g$ as the AAH modulation strength $W$, $2 \pi b$ as the twisting angle $\Delta \phi$, and $\beta$ as the AAH phase factor $\phi_\nu$. Thus, one can capture the essential physics of the AAH model using the above formulation through our helical system.

Before discussing the results, let us first mention that the present communication focuses on the right-handed helix. All the energies are measured in the units of eV. The on-site energies of the leads are taken to be zero. In the absence of any electric field, the on-site energies $\epsilon_n$ in the AFH are fixed to zero, and we choose the NNH strength $t_1 = 1\,$eV, and for the leads $t_0 = 2.5\,$eV. To work in the wide-band limit, we set $t_0>t_1$. The coupling strengths between the central region to the source and drain electrodes, characterized by the parameters $\tau_S$ and $\tau_D$, are fixed at $0.8\,$eV. For any other choices of parameter values, the physical picture qualitatively remains the same, which we confirm through our exhaustive numerical calculation.
 
\subsection{Energy eigenvalues and transmission spectra} 
 Let us begin the discussion with the spectral behavior of the antiferromagnetic helix in the presence of an electric field and also spin-dependent scattering parameter. Figure.~\ref{fig:LRH}(a) shows the eigenspectra of a typical short-range hopping AFH for $N=20$ sites with $h=0.5$, $v_g =1$, and $\beta = 0$ for both the up and down spins, shown by red and black colors, respectively. The spectra for the up and down spins are non-degenerate. Similarly, Fig.~\ref{fig:LRH}(b) shows the eigenspectra of long-range AFH keeping with parameters same as the SRH. The spectra are also non-degenerate for the two opposite spin cases. Moreover, each of the spectra is gapped and distributed in multiple bands. Thus, what we accumulate is that for both scenarios, we get non-zero spin separation in the presence of an electric field and spin-dependent scattering parameter. Here it is important to point out that, whenever we set the field strength to zero, so such separation among up and down spin energy eigenvalues takes place.  
 
The channel separation suggests that, a finite mismatch is expected between up and down spin transmission probabilities (which is the key requirement to have spin figure of merit). To reveal this fact, in Fig.~\ref{fig:3}, we plot spin-resolved transmission probabilities as function of energy. 
The transmission probabilities for up and down spin channels are shown by the red and black colors, respectively, for the SRH (Fig.~\ref{fig:3}(a)) and LRH (Fig.~\ref{fig:3}(b)) antiferromagnetic helices. The system size and the other parameters are considered the same as in Fig.~\ref{fig:LRH}. 

The up and down spin transmission spectra are different from each other both for the SRH and LRH helices. The transmission spectrum shows gapped~\cite{PhysRevLett.110.180403,PhysRevB.100.205402} nature for both the spin channels due to the presence of the electric field, which acts as a correlated to the system, as mentioned earlier. Now, in order to have a favorable spin TE response, the spin-resolved transmission spectrum must satisfy two conditions. First, the transmission spectrum should be asymmetric around a fixed energy~\cite{Mahan7436,CULLEN20102059}. Second, there must be a crossing between the up and down spin transmission spectrum. The first criterion is a general one, valid for both charge and spin TE cases, while the second one is desirable for a favorable spin TE response. In Figs.~\ref{fig:3}(a) and (b), a few of such crossovers in the transmission profile are marked with dotted ellipses with blue color and also shown in the insets for clarity. For example in Fig.~\ref{fig:3}(a), around the energy 1.4$\,$eV, on the left side of the crossing, there is a sharp peak in the down-spin transmission, while the up-spin transmission spectrum has a sharp peak on the right side. Such a sharp peak leads to an asymmetry of the transmission function and the crossings assure a large spin thermopower. We discuss this aspect in greater detail in the context of thermopower in the next sub-section.

\begin{figure*}[t!]
\centering
\includegraphics[width=0.325\textwidth,height=0.2\textwidth]{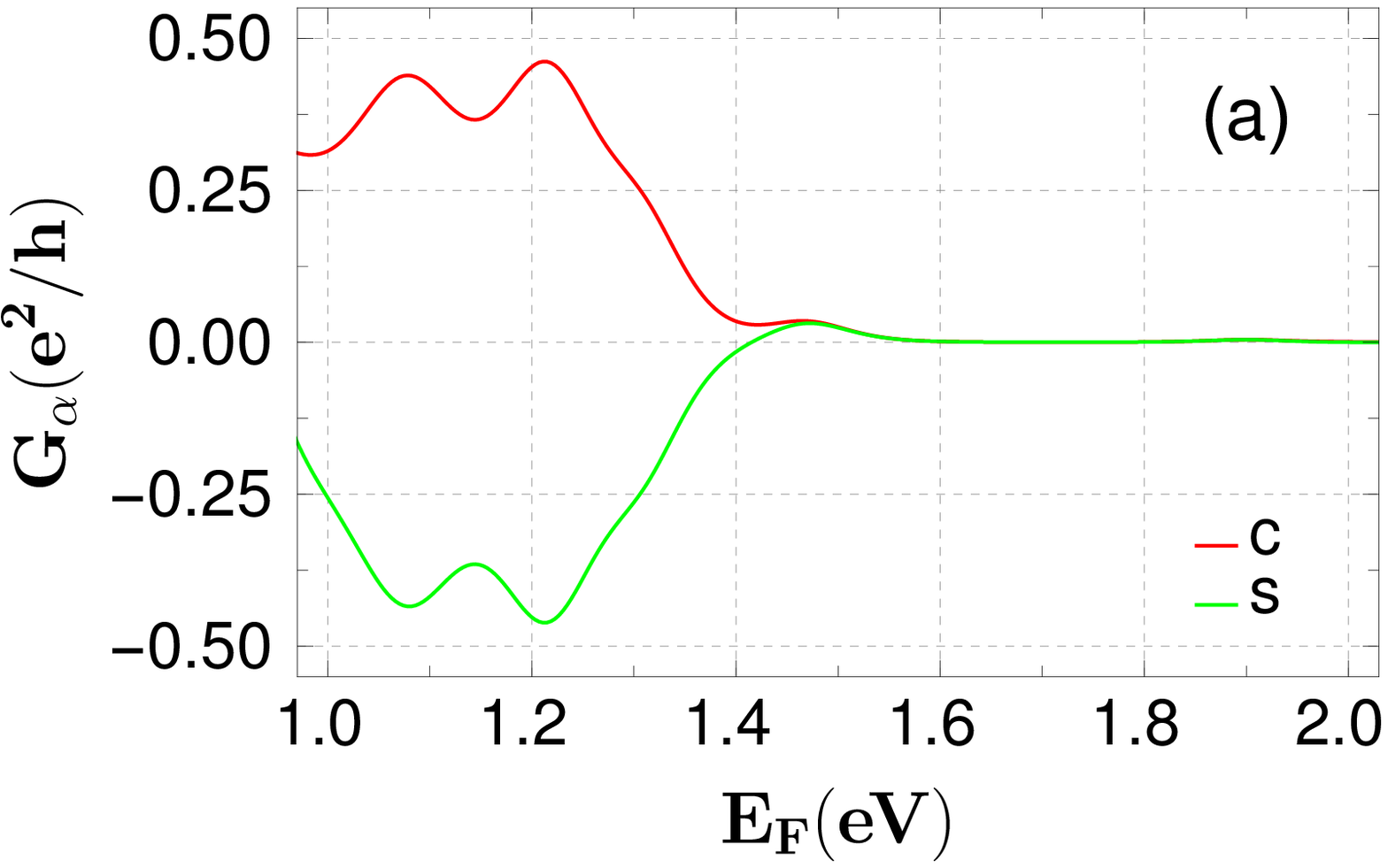} 
\includegraphics[width=0.325\textwidth,height=0.2\textwidth]{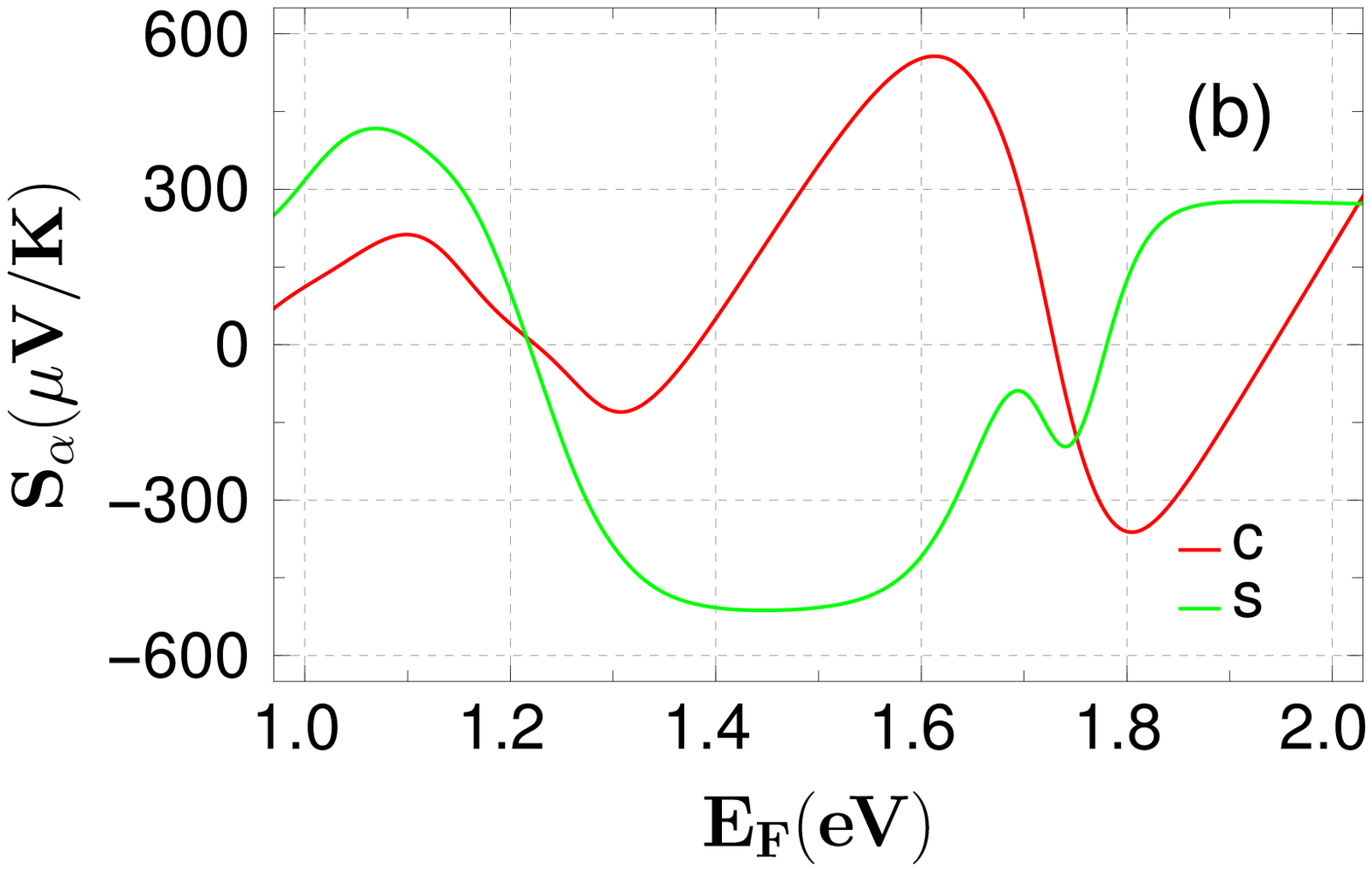} 
\includegraphics[width=0.325\textwidth,height=0.2\textwidth]{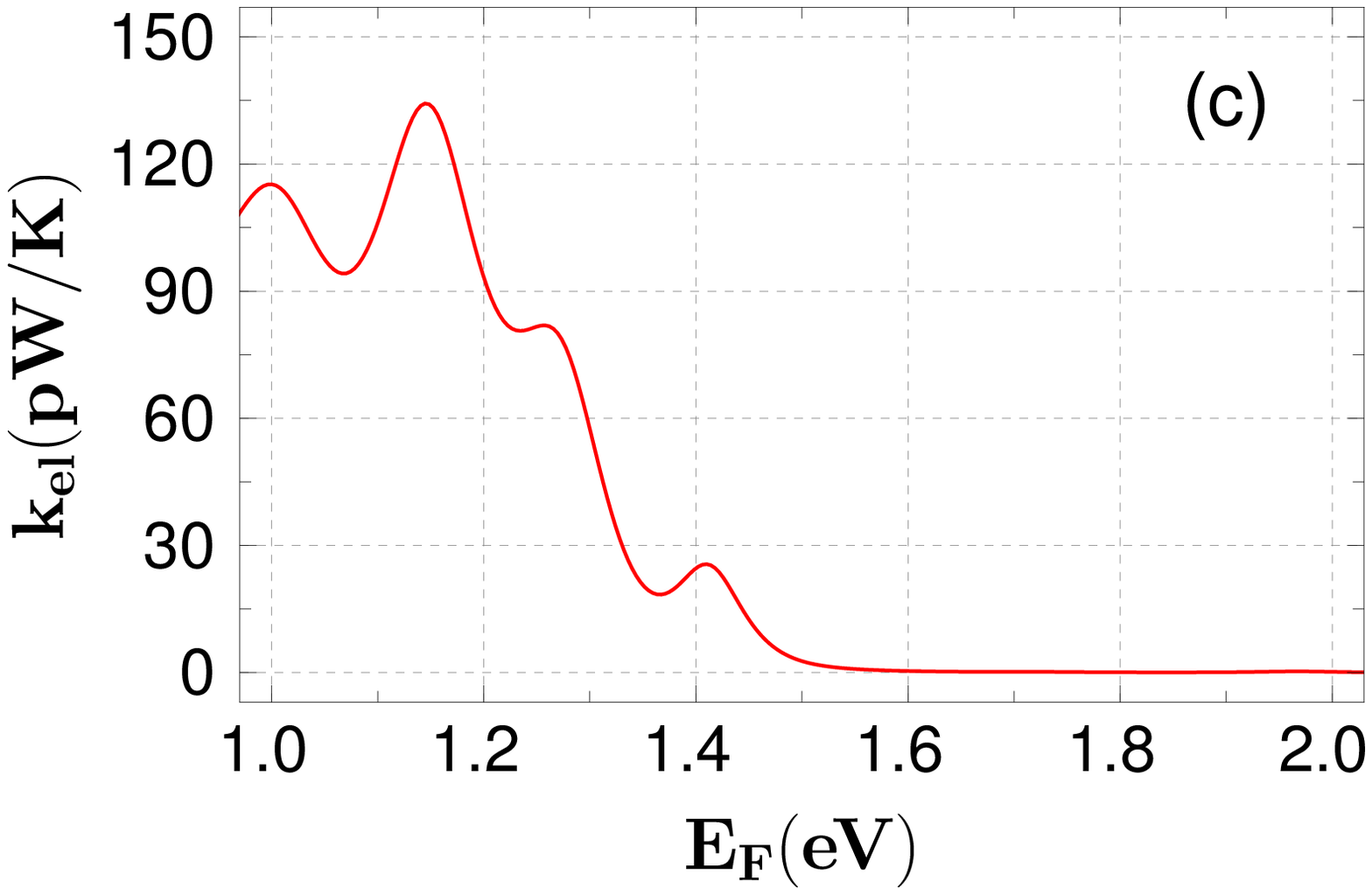} \vskip 0.1 in
\includegraphics[width=0.325\textwidth,height=0.2\textwidth]{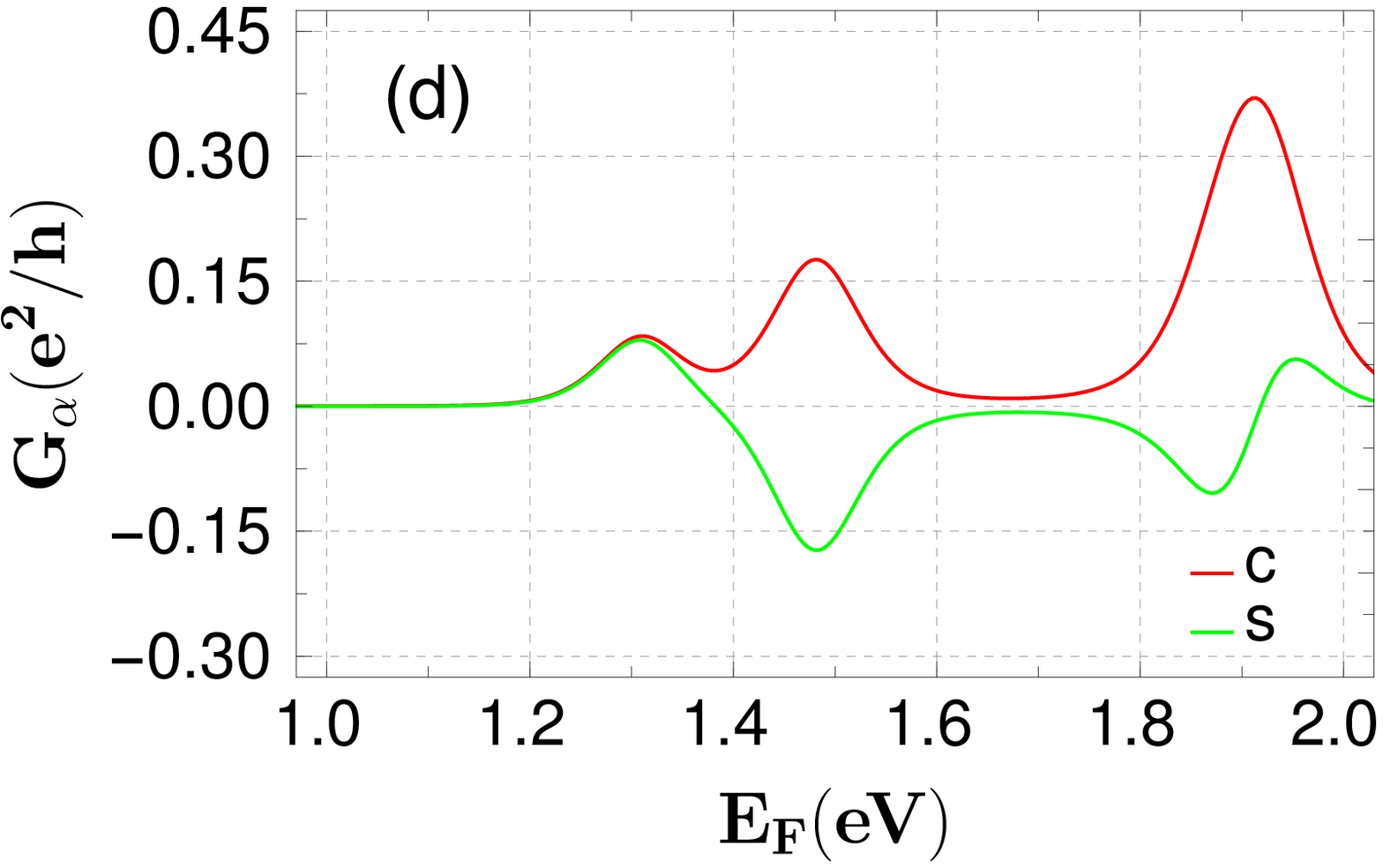} 
\includegraphics[width=0.325\textwidth,height=0.2\textwidth]{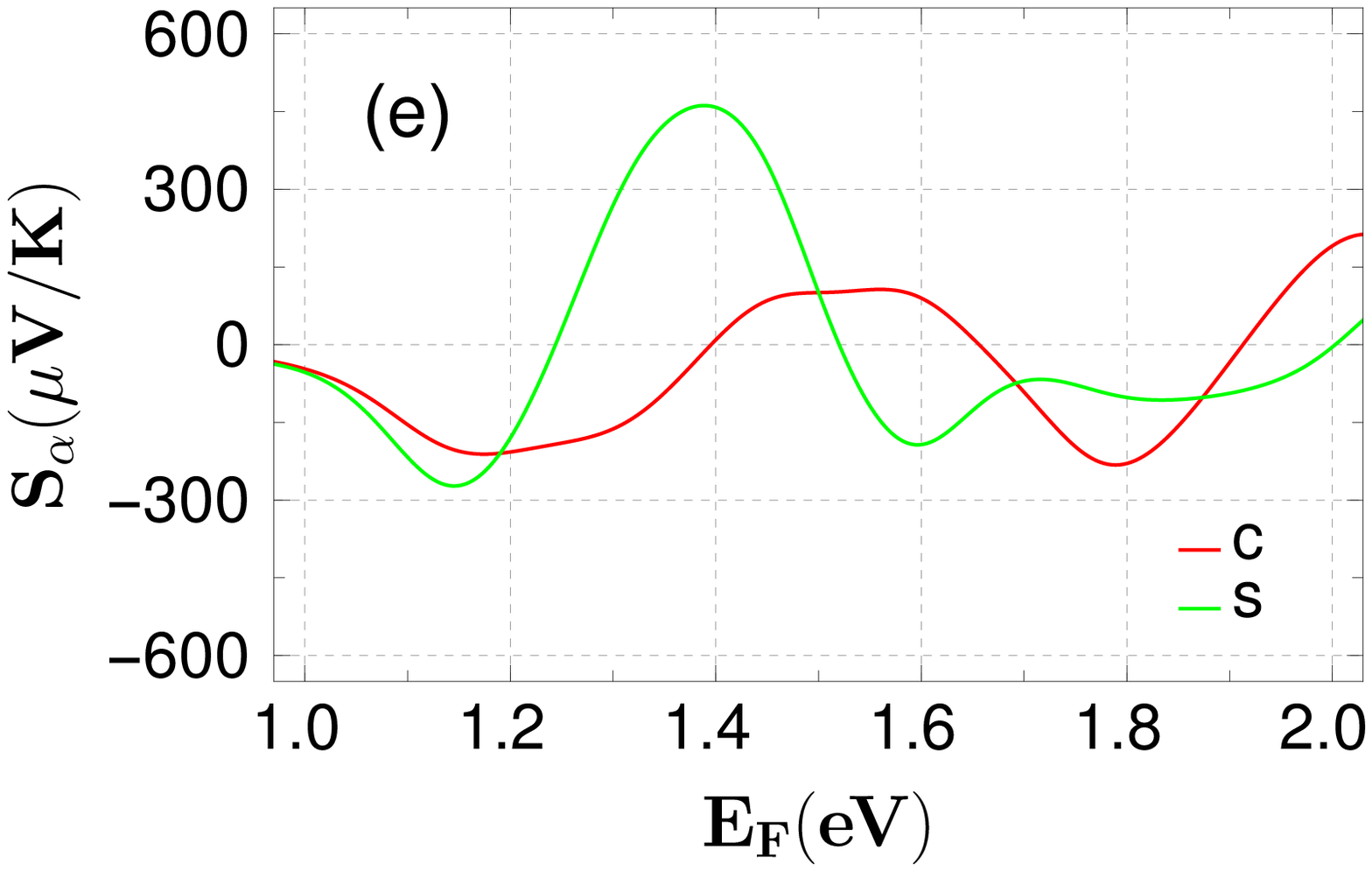}
\includegraphics[width=0.325\textwidth,height=0.2\textwidth]{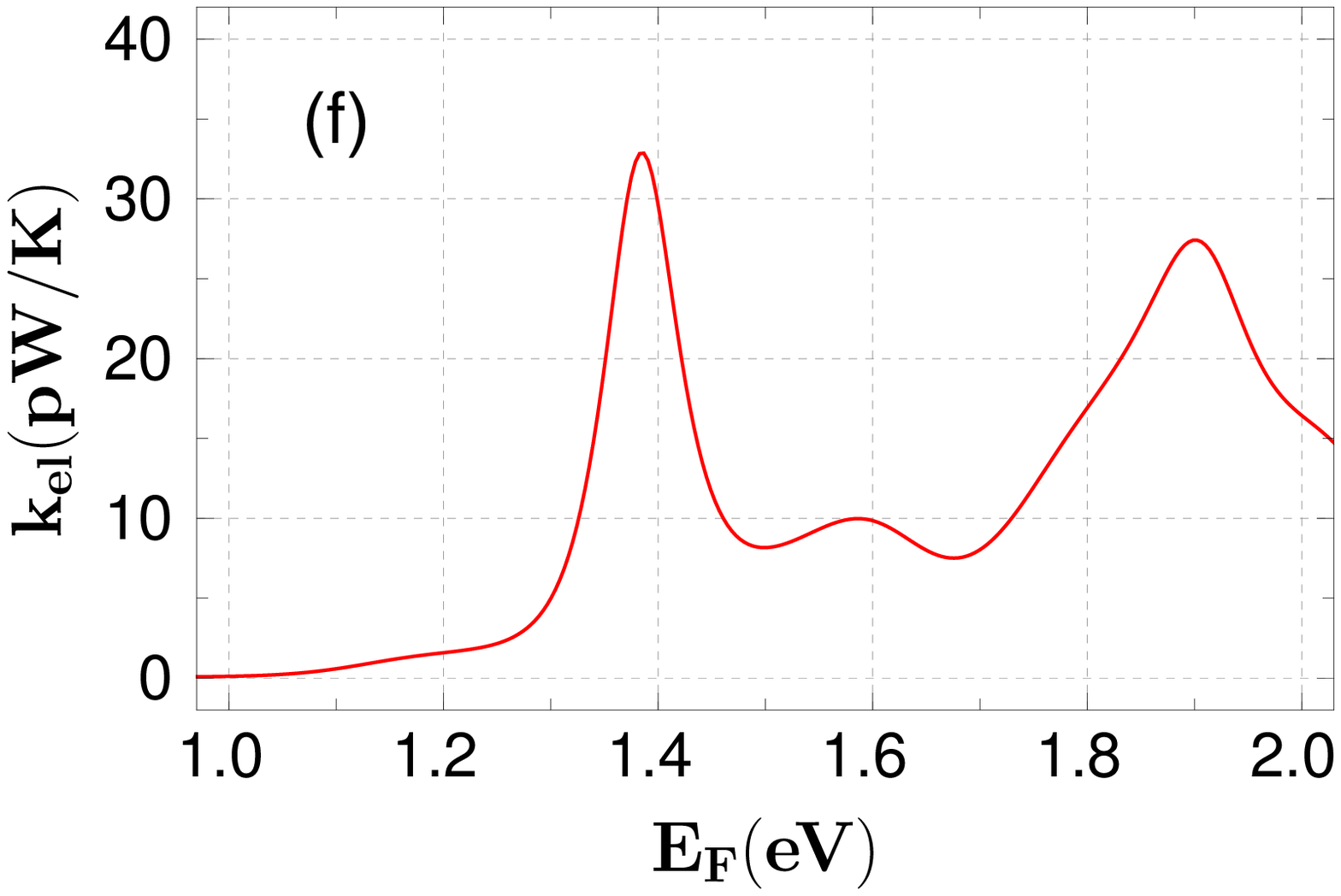}
\caption{(Color online). Behavior of different thermoelectric quantities at room temperature $T=300\,$K as a function of Fermi energy. The upper panel shows the results for the SRH helix and the lower panel for the LRH one. In (a) and (d) electrical conductance ($G_\alpha$), (b) and (e) thermopower ($S_\alpha$), and (c) and (e) thermal conductance due to electrons ($k_\text{el}$) are shown. All the parameters are the same as described in Fig.~\ref{fig:LRH}. The subscript $\alpha$ represents the charge ($c$) and spin ($s$) degrees of freedom and their corresponding results are shown by red and green curves, respectively.}
\label{fig:TEQ}
\end{figure*}


\subsection{Thermoelctric quantities}
Now, let us analyze the different TE quantities like electrical conductance, thermopower, thermal conductance, and figure of merit at room temperature ($T = 300 \,$K) in the presence of the electric field and spin-dependent scattering parameter. The charge and spin TE entities are computed for both the SRH and LRH helices.

The variation of electrical conductance $G_\alpha$ (in units of $e^2/h$) with Fermi energy $E_F$ is shown in Figs.~\ref{fig:TEQ}(a) and (b) for the SRH and LRH cases, respectively, where $\alpha$ corresponds to charge and spin. In Fig.~\ref{fig:TEQ}(a), we see that charge and spin dependent electrical conductances (shown by red and green curves, respectively) are almost symmetric about $G_\alpha=0$ line for the short-range helix. The maximum value of $|G_\alpha|$ is found to be $\approx 0.46$ and becomes vanishingly small beyond $E_F \sim 1.4\,$eV. This can be explained using the transmission profile of the SRH case. Remember, the total charge electrical conductance is defined as the sum of contributions coming from up and down spin channels, whereas the spin counterpart is the difference between the two. Since the charge and spin $G_\alpha$ are of the opposite signs below $E_F \sim 1.4\,$eV, it implies that the contribution from the up spin channel is vanishingly small. This is due to the fact that in this Fermi energy range, the up spin transmission probability is negligibly small compared to the down spin transmission probability as seen in Fig.~\ref{fig:3}(a). Now, in the range of Fermi energy from $\sim 1.4$ to $\sim 2$, we see that both the up and down transmission probabilities are small, leading to vanishingly small $G_\alpha$ as shown in Fig.~\ref{fig:TEQ}(a). The LRH helix for the same set of parameters shows somewhat similar behavior as observed in Fig.~~\ref{fig:TEQ}(d), reflecting the up and down spin transmission spectra shown earlier. 

\begin{figure*}[t!]
\centering
\includegraphics[width=0.475\textwidth]{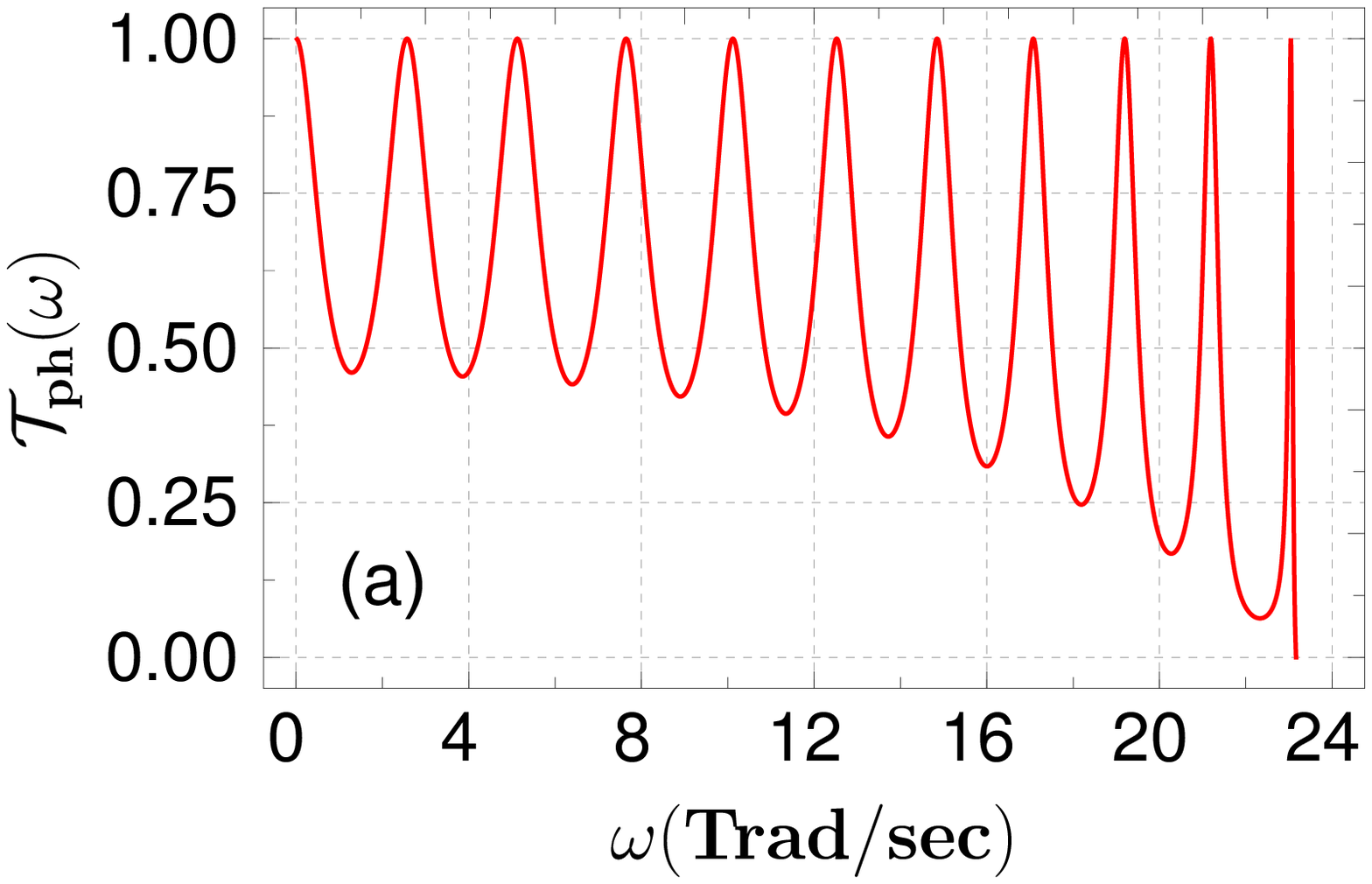} \hspace{0.2cm}
\includegraphics[width=0.455\textwidth]{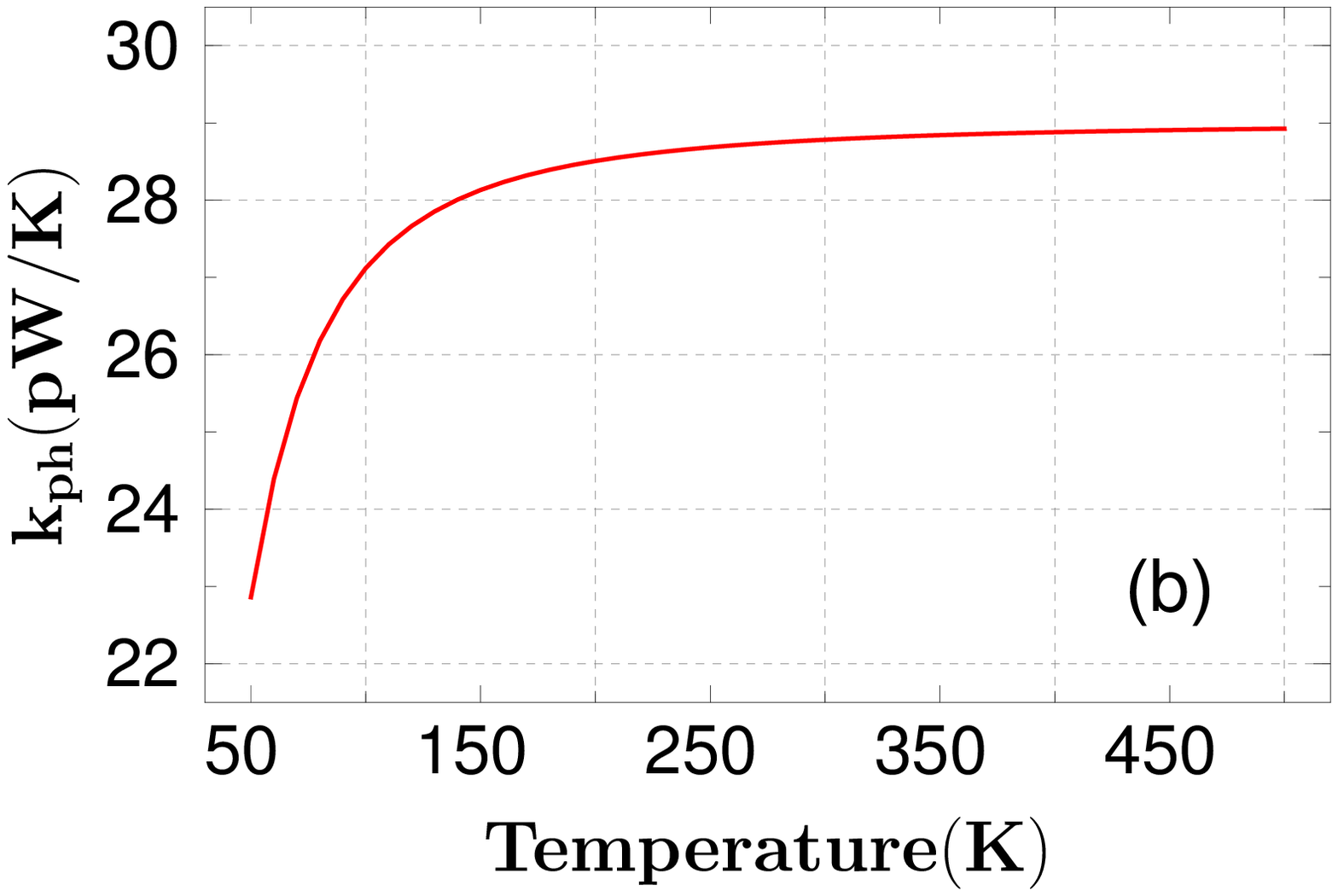} 
\caption{(Color online). (a) Phonon transmission probability $\mathcal{T}_\text{ph}$ as a function of phonon angular frequency $\omega$. (b) Phonon thermal conductance $k_\text{ph}$ as a function of temperature $T$.}
\label{fig:diagram5}
\end{figure*}


As the TE efficiency is directly proportional to the square of the thermopower, a large $S$ is always desirable. Moreover, in the case of spin TE, it is possible to achieve two different signs of thermopower, which can algebraically sum up to produce a larger value of $ZT$. Thus the choice of Fermi energy becomes very tricky. To have different signs of thermopower, one should look for a small region of the Fermi energy window where the transmission function is asymmetric. Not only that, the up and the down spin channels must have slopes of different signs. The thermopower is calculated using Eq. ~\ref{eq:TE-quantity}(b) and the corresponding Landaurer's integral $L_1$ where the transmission function is multiplied by $(E-E_F)$ and $\frac{\partial f_\text{FD}}{\partial E}$. The latter term provides thermal broadening and the product of the two is antisymmetric around the chosen $E_F$. As a result of that, if the transmission function is symmetric around $E_F$, then the thermopower will be zero irrespective of the value of the transmission probabilities. Now, if the slopes of the spin-resolved transmission functions are of the opposite signs around the chosen Fermi energy, then the thermopower picks a different sign with large values due to the asymmetric nature of the $\mathcal{T}(E)$. This will lead to a larger spin thermopower. Figurues~\ref{fig:TEQ}(b) and (e) show the variation of thermopower with Fermi energy in the same energy window as discussed for electrical conductance in the case of SRH and LRH, respectively. From the transmission profile of SRH helix (see Fig.~\ref{fig:3}(a)), it is clear that around $E\sim 1.4$, the up and down spin channels have a slope of different sign (shown by the blue dotted ellipse in Fig.~\ref{fig:3}), leading to a large value of spin thermopower as is seen in Fig.~\ref{fig:TEQ}(b). Since the charge thermopower is the algebraic sum of the up and down spins, respectively, it becomes very small at this Fermi energy. Similarly, one can explain the large value of the spin thermopower in case LRH as shown in Fig.~\ref{fig:TEQ}(e). Here too, the corresponding transmission profile shows that the up and down spin channels have a slope of opposite signs at $E_F \sim 1.38\,$eV, yielding large spin thermopower compared to its charge counterpart. The maximum thermopower is about 600$\,\mu$V/K for SRH helix and 550$\,\mu$V/K for the LRH one. 

The behavior of thermal conductance due to electrons as a function of Fermi energy is shown in Figs.~\ref{fig:TEQ}(c) and (f) for SRH and LRH helices, respectively. In the given Fermi energy window, the maximum value of thermal conductance is $\sim 135\,$pW/K (see Fig.~\ref{fig:TEQ}(c)) and is suppressed beyond $E_F\sim 1.5\,$eV. In the case of LRH, thermal conductance is found to have lower values than that for SRH. The maximum value is close to $33\,$pW/K around $E_F \sim 1.38\,$eV.

The system sizes considered in the present work are small (of the order of a few nm) and therefore it is expected that the thermal conductance due to phonons should have lower values compared to its electronic counterpart. However, for a precise estimation of the figure of merit, it is important to include the thermal conductance due to phonons, which we discuss now.
\subsection{Phonon contribution to thermal conductivity}

Before we discuss the behavior of $k_{\text{ph}}$, one needs to mention the spring constants of the electrodes and the central helix molecule. The 1D electrodes are considered Au electrodes, whose spring constant is $14.68\,$N/m~\cite{PhysRev.111.707}. For the helix molecule, we consider the spring constant about $5.1\,$N/m, which is considered as same as the single-crystal benzene~\cite{doi:10.1063/1.1725566}. Here we assume that two different atoms are adjacent to each other at the interface, one type of atom accounts for the Au electrode and the other type for the helix molecule. By averaging the spring constants of the electrodes and helix molecule, and the masses, the cut-off frequency for Au electrode comes out to be $\omega_c= 13.7\,$Trad/s. Here it should be noted that the spring constant for the helix molecule is chosen for a light molecule. However, if one works with heavy molecules, the phonon vibrations will be less than our case and therefore, $k_{\text{ph}}$ is expected to have lower values, and hence larger $ZT$.  

\begin{figure*}[t!]
\centering
\includegraphics[width=0.45\textwidth]{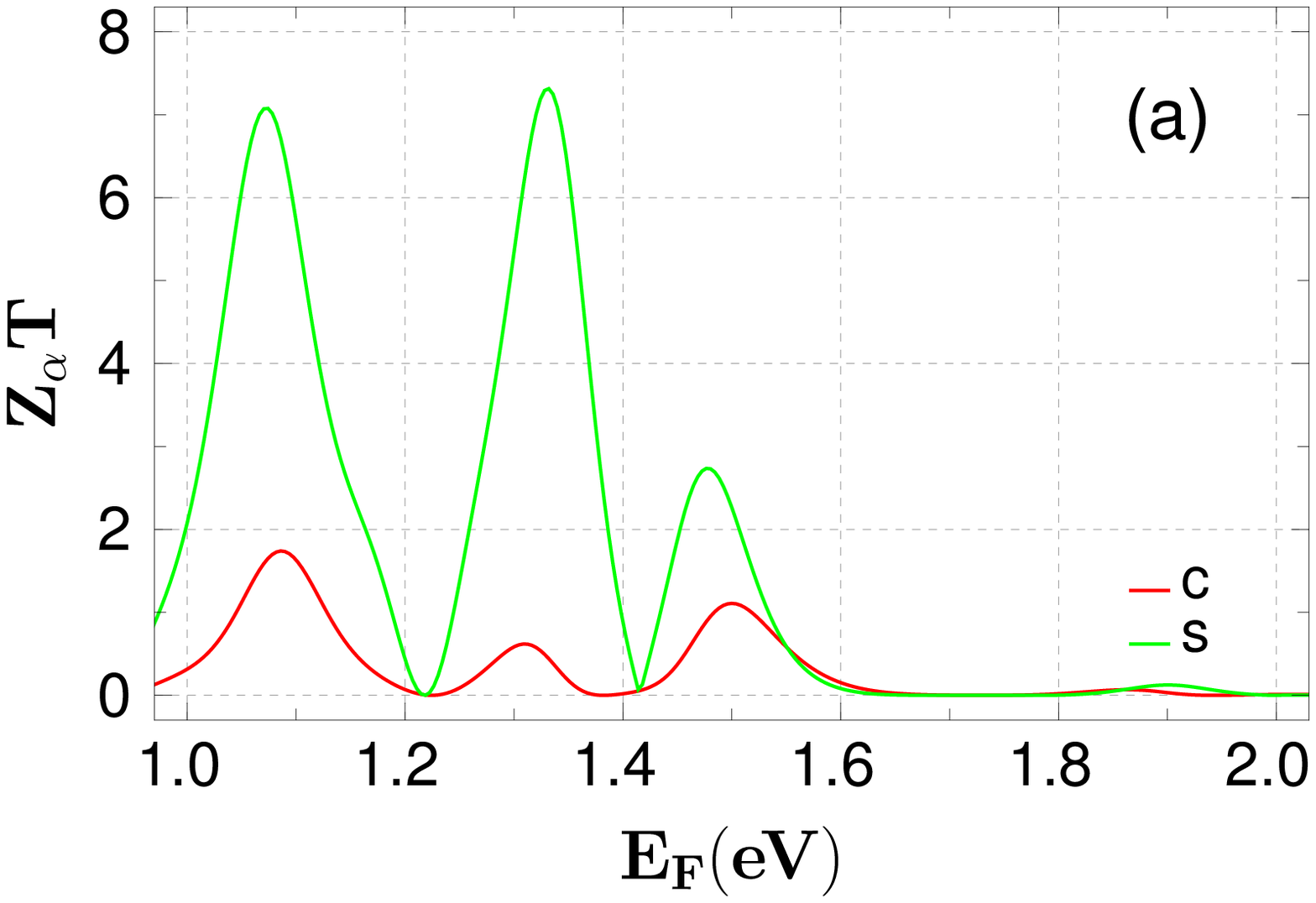} \hspace{0.2cm}
\includegraphics[width=0.465\textwidth]{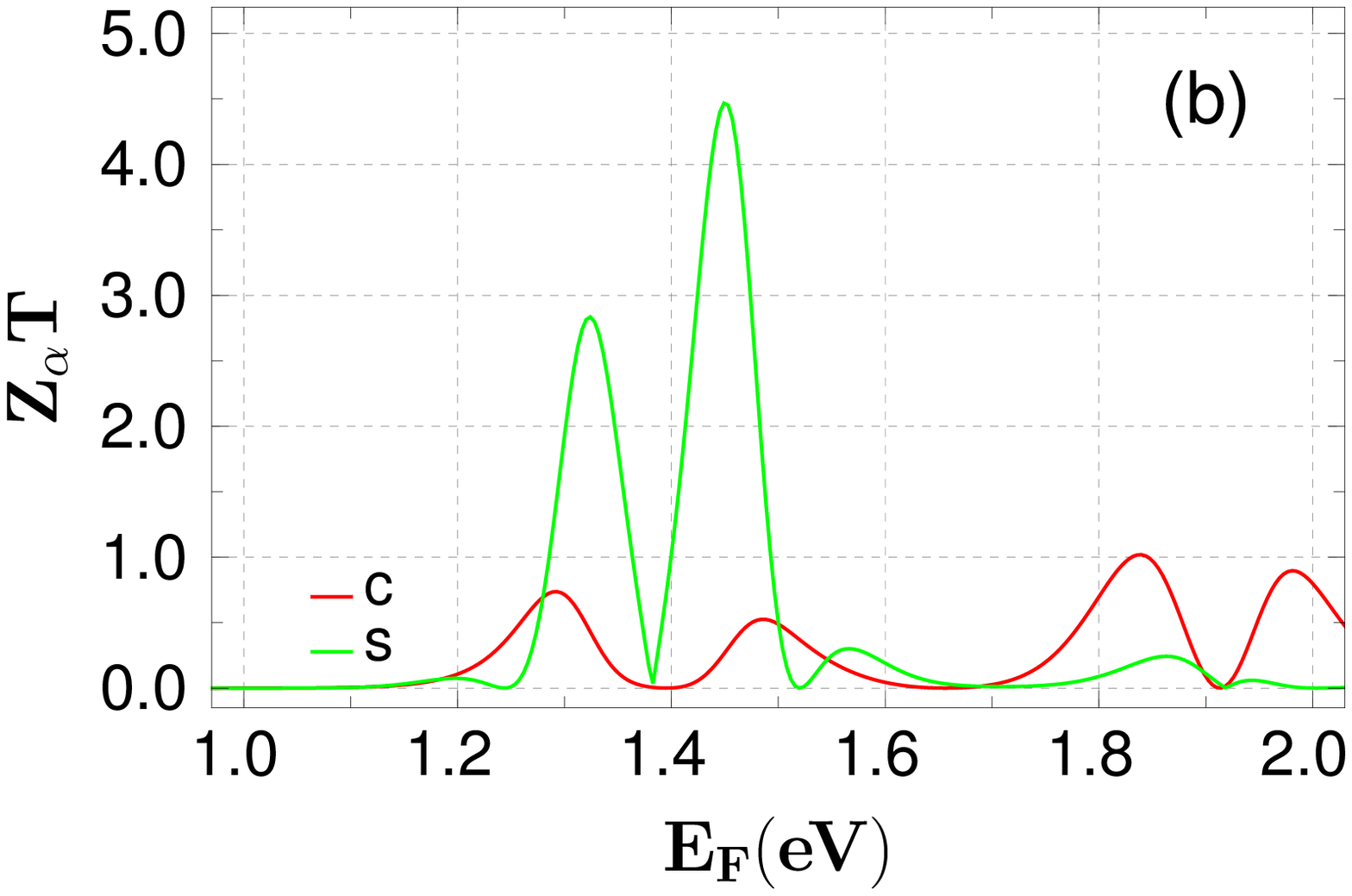} 
\caption{(Color online). Behavior of $Z_cT$ and $Z_sT$ as a function of Fermi energy at room temperature for (a) SRH  and (b) LRH helices. All the parameters are considered as the same as in Fig.~\ref{fig:LRH}. The red and green curves represent the results for charge and spin FOMs, respectively.}
\label{fig:diagram6}
\end{figure*}

In Fig.~\ref{fig:diagram5}(a), the phonon transmission probability is plotted as a function of phonon frequency. We observe a few Fabry-p\'{e}rot-like peaks~\cite{aghosh}. The behavior of phonon thermal conductance with temperature is shown in Fig.~\ref{fig:diagram5}(b). Within the temperature window 50 to 150$\,K$, $k_{\text{ph}}$ increases rapidly with temperature, and then it tends to saturate. The saturated value is about 29$\,$pW/K. 

\subsection{Thermoelectric efficiency}
With all the TE quantities and considering the phonon contribution, we finally compute FOM. At room temperature, the charge and spin $ZT$s  as a function of the Fermi energy are presented for SRH and LRH helices, respectively, as shown in Figs.~\ref{fig:diagram6}(a) and (b), respectively. Both SRH and LRH molecules exhibit favorable spin TE responses and dominate over their charge counterpart. Maximum spin-$ZT$ is obtained about $7$ for SRH and $4.5$ for LRH at $E_F \sim 1.35\,$eV and $E_F\sim 1.45\,$eV, respectively. Thus, our prescription indeed shows a favorable spin TE response at room temperature. 
\subsection{Role of $\beta$}
\begin{figure}[ht!]
\centering
\includegraphics[width=0.495\textwidth]{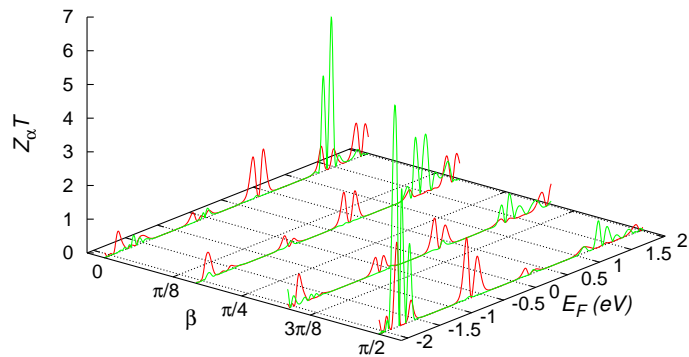} 
\caption{(Color online). Variation of charge and spin FOMs as a function of Fermi energy for different values of $\beta$ as shown by red and green colors, respectively for LRH helix. All the parameters are identical to those in Fig.~\ref{fig:LRH}. $\beta$-values are considered as $\beta=\pi/6$, $\pi/3$, and $\pi/2$. $\beta=0$ is included for comparison. }
\label{fig:diagram8}
\end{figure}

So far, the direction of the electric field was assumed to be parallel to the positive $\hat{x}$-axis, that is $\beta=0$. To study the effect of $\beta$ on TE performance, we consider other three different angles, namely, $\beta=\pi/6$, $\pi/3$, and $\pi/2$. The result for $\beta=0$ is also included for comparison. 
Figure.~\ref{fig:diagram8} shows the variation of charge and spin figure of merits in the case of LRH at room temperature as a function of Fermi energy for different values of $\beta$. All other parameters are kept fixed, as stated earlier. The variation of $Z_\alpha T$ as a function of Fermi energy varied from $-2.5$ to $3.5\,$eV, which is the full energy window as shown in Fig.~\ref{fig:LRH}. Mostly, in all the cases, spin-$ZT$ shows favorable response at different Fermi energies. Maximum spin-$ZT$ is noted about $4.5, 1.75, 0.8$, and $6.58$ for $\beta=0,\pi/6$, $\pi/3$, and $\pi/2$, respectively. Interestingly, maximum values of spin-$ZT$ dominate over the charge-$ZT$ for all the $\beta$ values considered here. The effect of $\beta$ on $Z_\alpha T$ for SRH will lead to more or less similar features and hence is not included for the brevity of the presentation.

\subsection{Effect of temperature}
\begin{figure}[ht!]
\centering
\includegraphics[width=0.495\textwidth]{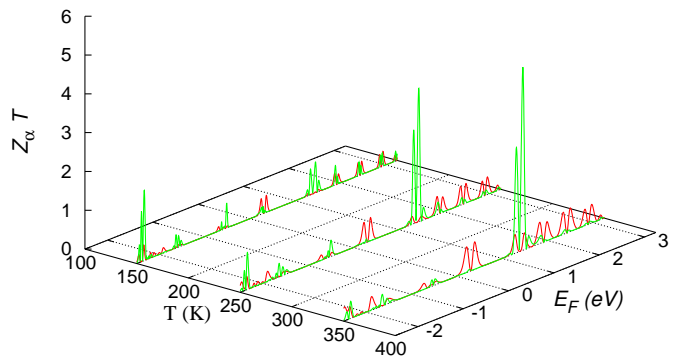} 
\caption{(Color online). Charge and spin thermoelectric FOMs as a function of Fermi energy at different temperatures as shown by the red and green colors, respectively for LRH helix. All the parameters are identical with Fig.~\ref{fig:LRH}.}
\label{fig:diagram7}
\end{figure}

All the results discussed so far are at room temperature $T=300\,$K. To study the effect of temperature, we have plotted $Z_\alpha T$ as a function of Fermi energy for three other different temperatures, namely, $T=150\,$K, 250$\,$K, and 350$\,$K, as shown in Fig.~\ref{fig:diagram7} for the LRH helix. The other parameters are kept fixed, as mentioned in Fig.~\ref{fig:3}. The spin $ZT$ and the charge $ZT$ are shown by the green and red colors, respectively. The temperature profile indicates that the maximum value of the spin figure of merit tends to increase with the increase in operating temperature. The maximum $Z_sT$, in this case, is found to be around $4.7$ at temperature $350\,$K.
 

\section{Conclusions}
In this present work, we have proposed a scheme to achieve a favorable spin TE response in a typical helical geometry with a spin configuration of antiferromagnetic texture. We have considered both the short and long-range hopping scenarios, which potentially mimic biological systems like single-stranded DNA and $\alpha$-protein molecules. We have considered the spin-dependent scattering phenomena and also a transverse electric field to study thermoelectric physics in the helical system.
In the absence of electric field or helicity, spin-dependent phenomena is no longer observed. We have used the NEGF formalism following the Landauer-Buttiker prescription to study the thermoelectric phenomena. Both the charge and spin TE responses have been studied. For a precise estimation of the TE {\it figure of merit}, we have computed the phonon contribution to the total thermal conductance. We have achieved a highly favorable spin TE response compared to the charge counterpart at room temperature for both the SRH and LRH molecules. The role of $\beta$, that is the angle between the direction of the electric field with the positive $\hat{X}$-axis on $ZT$ and the effect of temperature have also been examined. 

To the best of our concern, spin-dependent TE phenomena in antiferromagnetic helix has not been studied so far in the literature. Our proposition provides a new route of achieving efficient energy conversion using similar kinds of fascinating antiferromagnetic systems.

\bibliography{ref}
\bibliographystyle{apsrev4-1}

\end{document}